# Characteristics of plasmonic at a metal /chiral sculptured thin film interface

F. Babaei [a,*], M. Omidi [a]

[a)] Department of Physics, University of Qom, Qom, Iran

[*)] Email: fbabaei@qom.ac.ir

**Abstract**

The propagation of surface plasmon polariton at an interface of metallic thin film and chiral sculptured thin film theoretically has been investigated using the transfer matrix method in the Kretschman configuration. The optical absorption of structure as a function of polar incident angle for linear polarization P and S has been calculated at different structural parameters. The results show that exist multiple plasmon peaks for P-polarization, while there are the weak plasmon peaks when incident of light is S- polarized plane wave.

## 1. Introduction

Quasiparticles that can propagate at the interface between metal and dielectric due to collective charge oscillations are known as surface plasmon polaritons (SPP) [1]. Broad investigations of SPP waves have been done at the interface between a metal and an isotropic dielectric material [2, 3].The research was later extended to interfaces of metals and anisotropic dielectric materials [4, 5].

Since the introduction of sculptured thin films (STFs) in 1995 by Lakhtakia at a meeting held at Penn State, the study of SPP waves on the interface of a metal and sculptured dielectric thin film has been focused. The chiral sculptured thin films (CSTFs) are three dimensional bianisotropic nano-structures that can be produced by combination of oblique angle deposition (OAD) and rotation of the substrate about its surface normal [6-9].The existence of sharp peak in optical absorbance of structure as angle of incident of light with linear polarization is due to excitation surface plasmon-polariton wave [10].

A SPP is a surface electromagnetic wave, whose electromagnetic field is confined to the near vicinity of the dielectric–metal interface [11]. This confinement leads to an enhancement of the electromagnetic field at the interface, resulting in an extraordinary sensitivity of SPPs to surface conditions. The sensitivity of peak wavelengths of plasmonic properties between nanometal and dielectric medium has been investigated using the finite-element method[12-14].This sensitivity is extensively used for studying adsorbates on a surface, surface roughness, and related phenomena [11, 15].

In this study, we report on propagating SPP dependence to structural parameters of metal and CSTF. The theoretical formulation is outlined in Section 2. Numerical results are presented and discussed in Section 3.

## 2. Theory in brief

The Kretschmann configuration is a common experimental arrangement for the excitation and detection of SPP waves [2, 3]. In this study, the modified Kretschmann configuration was used to propagate of SPP waves [5]. The region $0 \leq z \leq l_{met}$ is occupied by a metal of relative permittivity $\varepsilon_{met}$, the region $l_{met} \leq z \leq (l_\Sigma = l_{met} + l_{STF})$ by a dielectric CSTF described by the angle of rise $\chi$ and structural of period (pitch) $2\Omega$, whereas the regions $z \leq 0$ and $z \geq l_\Sigma$ are homogeneous isotropic dielectric material of relative permittivity $\varepsilon_l = n_l^2$ (Fig.1). A plane wave in the half space $z \leq 0$ propagating at an angle $\theta_{inc}$ to the z- axis and at an angle $\psi_{inc}$ to the x- axis in the xy - plane towards the metal layer- coated CSTF in order to excite the SPP wave. The phasors of incident, reflected and transmitted electric fields are given as [16]:

$$\begin{cases} \underline{E}_{inc}(\underline{r})=[a_s\underline{S}+a_p\underline{P}_+]e^{i\underline{K}_0 n_l \cdot \underline{r}}, & z\leq 0 \\ \underline{E}_{ref}(\underline{r})=[r_s\underline{S}+r_p\underline{P}_-]e^{-i\underline{K}_0 n_l \cdot \underline{r}}, & z\leq 0 \\ \underline{E}_{tr}(\underline{r})=[t_s\underline{S}+t_p\underline{P}_+]e^{i\underline{K}_0 n_l \cdot \underline{r}}e^{i\underline{K}_0 n_l \cdot(\underline{r}-l_\Sigma \underline{u}_z)}, & z\geq l_\Sigma \end{cases} \quad (1)$$

The magnetic field's phasor in any region is given as:

$$\underline{H}(\underline{r}) = (i\omega\mu_0)^{-1}\underline{\nabla}\times\underline{E}(\underline{r})$$

where $(a_s, a_p)$, $(r_s, r_p)$ and $(t_s, t_p)$ are the amplitudes of incident plane wave, and reflected and transmitted waves with S- or P- polarizations. We also have:

$$\begin{cases} \underline{r}=x\underline{u}_x+y\underline{u}_y+z\underline{u}_z \\ \underline{K}_0 = K_0(\sin\theta_{inc}\cos\psi_{inc}\underline{u}_x + \sin\theta_{inc}\sin\psi_{inc}\underline{u}_y + \cos\theta_{inc}\underline{u}_z) \end{cases} \quad (2)$$

where $K_0 = \omega\sqrt{\mu_0\varepsilon_0} = 2\pi/\lambda_0$ is the free space wave number, $\lambda_0$ is the free space wavelength, $\varepsilon_0 = 8.854\times10^{-12}\,Fm^{-1}$ and $\mu_0 = 4\pi\times10^{-7}\,Hm^{-1}$ are the permittivity and permeability of free space (vacuum), respectively. The unit vectors for linear polarization normal and parallel to the incident plane, $\underline{S}$ and $\underline{P}$, respectively are defined as:

$$\begin{cases} \underline{S} = -\sin\psi_{inc}\underline{u}_x + \cos\psi_{inc}\underline{u}_y \\ \underline{P}_\pm = \mp(\cos\theta_{inc}\cos\psi_{inc}\underline{u}_x + \cos\theta_{inc}\sin\psi_{inc}\underline{u}_y) + \sin\theta_{inc}\underline{u}_z \end{cases} \quad (3)$$

and $\underline{u}_{x,y,z}$ are the unit vectors in Cartesian coordinates system.

The reflectance and transmittance amplitudes can be obtained, using the continuity of the tangential components of electrical and magnetic fields at interfaces and solving the algebraic matrix equation [5]:

$$\begin{bmatrix} t_s \\ t_p \\ 0 \\ 0 \end{bmatrix} = [\underline{\underline{K}}]^{-1}\cdot[\underline{\underline{B}}]\cdot[\underline{\underline{M}}'_{CSTF}]\cdot e^{i[\underline{\underline{P}}_{met}]l_{met}}\cdot[\underline{\underline{K}}]\cdot\begin{bmatrix} a_s \\ a_p \\ r_s \\ r_p \end{bmatrix} \quad (4)$$

The different terms and parameters of this equation are given in detail by Lakhtakia [16]. In order to obtain transfer matrix ($[\underline{\underline{M}}'_{CSTF}]$) the piecewise homogeneity approximation method [17, 18] is used. In this method, the CSTF is divided into N sub layers with a thickness of $h = d/N$ (2 nm will suffice). The reflection and transmission can be calculated as:

$$R_{i,j} = \left|\frac{r_i}{a_j}\right|, \quad T_{i,j} = \left|\frac{t_i}{a_j}\right|^2 \quad ; \quad i,j = s,p \tag{5}$$

### 3. Numerical results and discussion

For the purpose of simulation, the structure is considered as a right-handed MgF$_2$ CSTF and silver thin film. In optical modeling, the relative permittivity scalars $\varepsilon_{a,b,c}$ of CSTF were calculated using the Bruggeman homogenization formalism [19], as it is shown in Fig.2. In this formalism, the structure is considered as a two component composite (MgF$_2$ and void). These quantities are dependent on different parameters, namely, columnar form factor, fraction of MgF$_2$ ($f_{MgF_2}$), the wavelength of free space and the refractive index $n(\lambda_0) + ik(\lambda_0)$. In addition, each column in the structure is considered as a string of identical long ellipsoids [17]. The ellipsoids are considered to be electrically small (i.e. small in a sense that their electrical interaction can be ignored) [20]. In all calculations columnar form factors $(\frac{c}{a})_{MgF_2} = (\frac{c}{a})_{void} = 20$, $(\frac{b}{a})_{MgF_2} = (\frac{b}{a})_{void} = 1.06$ (c, a, b are semi major axis and small half-axes of ellipsoids, respectively) [21] .We have used the bulk experimental refractive indexes silver and MgF$_2$ [22] (Fig.2). In all our calculations, the absorbances $A_i = 1 - \sum_{j=s,p} R_{ji} + T_{ji}$, $i = s,p$ as a function of $\theta_{inc}$ are calculated when $\lambda_0 \cong 633nm$ (a Helium-

Neon laser with a power of 12 mW operating at $\lambda_0 = 632.8\,nm$ was considered as a source of monochromatic, collimated light)(Figs. 3-8).

The calculated optical absorption of structure as a function of $\theta_{inc}$, when $\psi_{inc} = 0°$, $\lambda_0 = 633\,nm$, at different thicknesses of sliver for S- and P- polarized plane wave is plotted in Fig.3. The CSTF (MgF$_2$) is described by the following parameters: $f_v = 0.4$, $\chi = 20°$, $\Omega = 150\,nm$ and $l_{CSTF} = 2\Omega$. The relative permittivity of the silver (Ag) is $\varepsilon_{Ag} = -15.9761 + i\,1.03309$, the ambient medium is $\varepsilon_l = 4$ (silicon nitride ($Si_3N_4$)). Fig.3 shows that as the thickness of metal increases the SPP peak shifts to shorter polar incident angles for P- polarized plan wave. However, at first the intensity of absorption increases (close to the skin depth $\delta_{met} = \dfrac{\lambda_\circ}{2\pi\,\text{Im}[\sqrt{\varepsilon_{met}}]}$ of metal) and then decreases. The skin depth of silver at $\lambda_0 = 633\,nm$ is 25.21nm. The intensity of absorption can be increased by choosing a proper void fraction and higher structural of period of CSTF. If thickness of metal more than skin depth, before the incident of light to interface of metal and CSTF, is absorbed and or reflected. Because, with the increase of the metal film thickness the efficiency of the SPP excitation (and the field enhancement) decreases as the tunneling distance increases [11].If thickness of metal less than skin depth, a high percentage of incident light is transmitted of metal thin film and SPP is not excited. Therefore, when thickness of metal is close to the skin depth, the intensity of SPP is significant. The calculations were repeated for S-polarization plane wave and the same results obtained. However, in all our calculations the SPP peaks are weak, because the direction of the polarization is perpendicular to the plane of incidence and can be increased by suitable structure [23].

The optical absorption $A_P$ and $A_S$ as a function of $\theta_{inc}$, when $l_{met} = 30nm$ at different porosities of CSTF have been depicted in Fig.4 (see caption of Fig.3 for structural parameters). Fig.4 shows that as the porosity of CSTF increases the SPP peak shifts to shorter polar incident angles for P- polarized plan wave and also its intensity decreases. At $f_v = 0$ the CSTF is homogeneous, isotropic and dense. Then the relative permittivity scalars $\varepsilon_{a,b,c}$ of CSTF at higher porosities decrease. Therefore the SPP peak roughly occurs close to critical angle $\sin\theta_{SPP} \cong \sqrt{\frac{\max(\varepsilon_a, \varepsilon_b, \varepsilon_c)}{\varepsilon_l}}$ [16] and it shifts to lower polar angles when porosity of CSTF increases. Also, the SPP peak widens, becomes asymmetric and eventually decreases as the porosity increases [24].

The calculations optical absorption as a function of $\theta_{inc}$ respectively were repeated, when $f_v = 0.4$ at different thicknesses of CSTF (fig.5), when $l_{CSTF} = 2\Omega$ at different pitches of CSTF (fig.6), when $\Omega = 150nm$ at different angles of rise of CSTF (fig.7) and when $\chi = 20°$ at different $\psi_{inc}$s incident light (fig.8) (see captions of them for structural parameters).

Fig.5 shows in our work at first the intensity of SPP peak increases until $l_{CSTF} = 6\Omega$, then is not changed and also $\theta_{SPP} \approx 40°$ is fixed. Calculations show that, as the normalized $l_{CSTF}/\Omega$ increases from 0, the intensity of SPP peak rises. This peak saturates, further increases of $l_{CSTF}/\Omega$ have imperceptible effects. As the structural of period of CSTF increases the intensity of SPP peak rises and shifts to shorter polar incident angles (fig.6).Because the thickness of thin film increases with increasing the structural of CSTF. The location and intensity of SPP is almost fixed as the angle of rise of CSTF increases (fig.7). In the growth of CSTFs, the angle of rise and film porosity can be controlled independently [25], and then the optical constants of thin film do not change with the angle of rise. Fig.8 shows that the optical absorption is not dependent on azimuthal angle, because in our work the cross section columns of CSTF is circular in xy plane (the small half-axes of ellipsoids are selected as $(\frac{b}{a})_{MgF_2} = (\frac{b}{a})_{void} \cong 1$) and $\varepsilon_a$'s behavior is similar to that of $\varepsilon_c$ ($\varepsilon_a \cong \varepsilon_c$).

In Fig.9 the effects of dispersion and dissipation of CSTF and metal on optical absorption spectra are considered. In the dispersion curve, the dispersion of dielectric function of CSTF is included in the Bruggeman homogenization formalism (i.e. homogenization is implemented for each wavelength). In other three plots, the homogenization is performed at fixed wavelength. In the dispersion curves, the relative permittivity scalars, $\varepsilon_{a,b,c}$ for each $\lambda_0$ are different. But in other curves, the relative permittivity scalars for each wavelength remain constant with $\varepsilon_{a,b,c}$ wavelength of homogenization [26]. On the other hand, these effects are significant on propagation SPP wave. Relative phase speed $K_\circ / K_{SPP}$ of SPP wave modes ($K_{SPP} = K_0 n_l \sin\theta_{SPP}$) with respect to the inverse-periodicity parameter $\Omega_\circ / \Omega$ ($\Omega_0$ is normalized pitch) for a P polarized plane wave has been calculated, as shown in Fig.10. It can be seen that there exist five plasmonic modes. In the limit $\Omega \to \infty$, only one mode is possible, because a chiral sculptured thin film is a columnar thin film [5] and it can be observed of Fig.10. Although, at higher structural periods due to numerical instabilities (overflow/underflow) occurred in the computation, could not find a fixed point of relative phase speed of plasmonic mode, but one can find its from extrapolate the resulting curve to the axis where $\Omega_\circ / \Omega = 0$. We found that four other SPP modes arise and then vanish as $\Omega \to 0$ [5]. At first mode 5 vanishes, followed modes 4, 3 and 2, respectively. Mode 1 alone survives as half structural period is reduced further and the results achieved in this work are consistent with Lakhtakia's group [5, 16].

In order to proof of excitation SPP wave at metal/CSTF interface, the time-averaged Poynting vector $P(z) = \frac{1}{2}\text{Re}[E(z) \times H^*(z)]$ versus thickness of metal film for a P polarized plane wave has been depicted in Fig.11. The compare between Cartesian components of P(z) shows that as metal thin film thickness increases the energy of photons of incident light is transferred to SPP quasiparticles and the SPP wave localized at metal/CSTF interface.

## 4. Summary

In this research, the excitation of SPP at the interface of silver thin film and chiral sculptured $MgF_2$ thin film, using the transfer matrix method in the Kretschman configuration theoretically has been investigated. The optical absorption of structure for linear polarization S and P as a function of incident angle has been studied. The effects of different structural parameters and dispersion dielectric function on the propagation SPP wave have been calculated. The calculations of optical modeling show that for linear polarization P, multiple SPP mode exist at the interface which propagate with different velocities. However, for linear polarization S, the structure shows weak plasmon peak. Therefore, this study may be applied to characteristic for plasmonic properties at the interface metallic structures and chiral anisotropic thin films.

**Acknowledgements**

We wish to acknowledge support from the University of Qom.

# References

1. Karpinski P, Miniewicz A (2011) Surface Plasmon Polariton Excitation in Metallic Layer Via Surface Relief Gratings in Photoactive Polymer Studied by the Finite-Difference Time-Domain Method. Plasmonics 6:541–546
2. Kretschmann E, Raether H, (1968) Radiative decay of nonradiative surface plasmons excited by light. Z. Naturforsch. A 23: 2135-2136
3. Simon H J, Michell D E, Watson J G (1975) Surface plasmons in silver films- a novel undergraduate experiment. Am. J. Phys 43: 630-636
4. Mihalache D, Baboiu D M, Ciumac M, Torner L, Torres J P(1994)Hybrid surface plasmon polaritons guided by ultrathin metal films .Opt. Quant. Electron. 26: 857-863
5. Polo Jr J A, Lakhtakia A(2009) On the surface plasmon polariton wave at the planar interface of a metal and a chiral sculptured thin film. Proc. R. Soc. A456: 87-107
6. Robbie K, Brett M J, Lakhtakia A (1996) Chiral sculptured thin films. Nature 384 : 616
7. Savaloni H, Babaei F, Song S, Placido F (2011) Influence of substrate rotation speed on the nanostructure of sculptured Cu thin film. Vacuum:85(2011)776-781
8. Savaloni H, Babaei F, Song S, Placido F (2009) Characteristics of sculptured Cu thin films and their optical properties as a function of deposition rate. Appl. Surf. Sci 55 :8041-8047
9. Song S, Keating M, Chen Y , Placido F(2012) Reflectance and surface enhanced Raman scattering (SERS) of sculptured silver films deposited at various vapor incident angles. Meas. Sci. Technol 23 : 084007- 084012
10. Mansuripur M, Zakharian A R, Moloney J V(2007) Surface plasmon polaritons on metallic surfaces .Opt. Photon. News 18: 44-49
11. Zayats A V, Smolyaninov I I, Maradudin A A (2005) Nano-optics of surface plasmon polaritons. Physics Reports 408 : 131–314
12. Yuan-Fong Chau, Zheng-Hong Jiang(2011)Plasmonics Effects of Nanometal Embedded in a Dielectric Substrate. Plasmonics 6:581-589
13. Yuan-Fong Chau(2012) Long-ranging propagation based on resonant coupling effects using a series connection of ten nanoshells in a plasmon waveguide. Applied optics 51: 640–643
14. Yuan-Fong Chau, Han-Hsuan Yeh, Din Ping Tsai (2008 Near-field optical properties and surface plasmon effects generated by a dielectric hole in a silver-shell nanocylinder pair. Applied optics 47: 5557–5561
15. Pitarke J M, Silkin V M, Chulko V E V,Echenique P M(2007) Theory of surface plasmons and surface-plasmon Polaritons. Rep. Prog. Phys 70 :1–87
16. Lakhtakia A (2007) Surface-plasmon wave at the planar interface of a metal film and a structurally chiral medium.Optics Commun 279:291-297
17. Lakhtakia A, Messier R (2005) Sculptured thin films, Nanoengineered Morphology and Optics. SPIE, USA
18. Polo Jr J A, Lakhtakia A(2004) Comparison of two methods for oblique propagation in helicoidal bianisotropic mediums. Opt. Commun 230:369-386
19. Lakhtakia A(2000) On percolation and circular Bragg phenomenon in metallic, helicoidally periodic, sculptured thin films. Microwave and Optical Technology Letters 24 : 239-244

20. Babaei F, Savaloni H (2008) Numerical study of the remittances of axially excited chiral sculptured zirconia thin films. Journal of Modern Optics 55 :1845-1857
21. Sherwin J A, Lakhtakia A, Hodgkinson I J (2002) On calibration of a nominal structure - property relationship model for chiral sculptured thin films by axial transmittance measurements, Opt. Commun 209 : 369- 375
22. Palik E D(1985)Handbook of Optical Constants of Solids, Academic Press, New York,USA
23. Gilani T H, Dushkina N, Freeman W L, Numan M Z, Talwar D N, Pulsifer D P(2010) Surface plasmon resonance due to the interface of a metal and a chiral sculptured thin film. Optical Engineering 49: 120503-1-120503-3
24. Abdulhalim I, Lakhtakia A, Lahav A, Zhang F, Xu J (2008) Porosity Effect on Surface Plasmon Resonance from Metallic Sculptured Thin Films. Proc. SPIE 7041: 70410C
25. Robbie K, Sit J C, Brett M J (1998) Advanced techniques for glancing angle deposition. J. Vac. Sci. Technol. B 16:1115-1122
26. Babaei F (2011) Effects of dispersion and dissipation on the optical rotation and ellipticity of chiral sculptured thin films .J. Mod. Optics 58: 1292-1296

**Figure captions**

Fig.1. Schematic of the boundary- value problem for SPP propagation.

Fig.2. a) The dielectric constants of bulk silver and magnesium fluoride, b ) the relative permittivity scalars $\varepsilon_{a,b,c}$ of CSTF as a function of porosity at $\lambda_0 = 633 nm$, c) the relative permittivity scalars $\varepsilon_{a,b,c}$ of CSTF as a function of wavelength at $f_v = 0.4$.

Fig . 3 . Calculated absorbance as a function of $\theta_{inc}$ , when $\psi_{inc} = 0°, \lambda_0 = 633 nm$, at different thicknesses of sliver for S- and P- polarized plane wave. The CSTF (MgF$_2$) is described by the following parameters: $f_v = 0.4, \chi = 20°, \Omega = 150 nm$ and $l_{CSTF} = 2\Omega$. The relative permittivity of the silver (Ag) is $\varepsilon_{Ag} = -15.9761 + i\,1.03309$, the ambient medium is $\varepsilon_l = 4$ (silicon nitride ($Si_3N_4$)).

Fig.4. Same as Fig. 3 except that the absorbance $A_P$ and $A_S$ were calculated at different porosities of CSTF and $l_{met} = 30nm$.

Fig.5. Same as Fig. 4, except that the absorbance $A_P$ and $A_S$ were calculated at different thicknesses of CSTF and $f_v = 0.4$.

Fig.6. Same as Fig. 5, except that the absorbance $A_P$ and $A_S$ were calculated at different half structural periods of CSTF and $l_{CSTF} = 2\Omega$.

Fig.7. Same as Fig. 6, except that the absorbance $A_P$ and $A_S$ were calculated at different angles of rise of CSTF and $\Omega = 150 nm$.

Fig . 8. Same as Fig. 7, except that the absorbance $A_P$ and $A_S$ were calculated at different $\psi_{inc}$s and $\chi = 20°$.

Fig.9. Same as Fig. 3, except that the absorbance $A_P$ and $A_S$ were plotted as a function of wavelength at homogenization wavelengths 553,633,733nm and [400-800nm], $l_{CSTF} = 4\Omega$, $l_{met} = 30 nm$, $\theta_{inc} = 40.3°$ and $f_v = 0.4$.

Fig.10. Relative phase speed $K_°/K_{SPP}$ of SPP wave modes versus the inverse-periodicity parameter $\Omega_°/\Omega$, when $\psi_{inc} = 0°, \lambda_0 = 633 nm$. The CSTF (MgF$_2$) is described by the following parameters: $f_v = 0.2$, $\chi = 35°$, $\Omega_° = 150 nm$, $l_{met} = 30 nm$ and $l_{CSTF} = 4\Omega$.

Fig.11. Cartesian components of the time-averaged Poynting vector P(z) in the metal film vs. $z \in (0, L_{met})$ when a surface-plasmon wave has been excited ($\psi_{inc} = 0°, \lambda_0 = 633 nm$, $\theta_{inc} = 46°$). The CSTF (MgF$_2$) is described by the following parameters: $f_v = 0.2$, $\chi = 35°$, $\Omega = 150 nm$, $l_{met} = 30 nm$ and $l_{CSTF} = 4\Omega$.

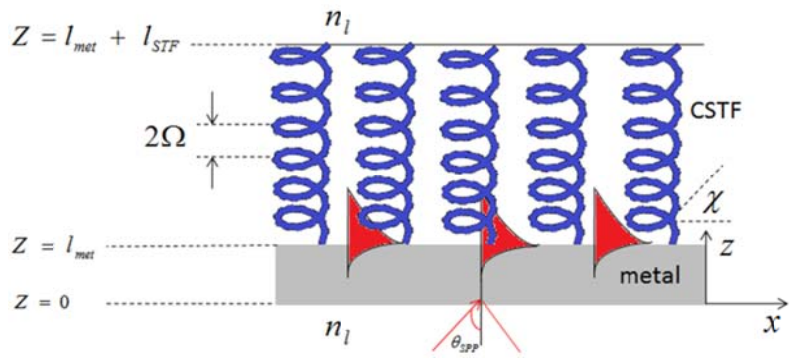

**Fig. 1;   F. Babaei and M. Omidi**

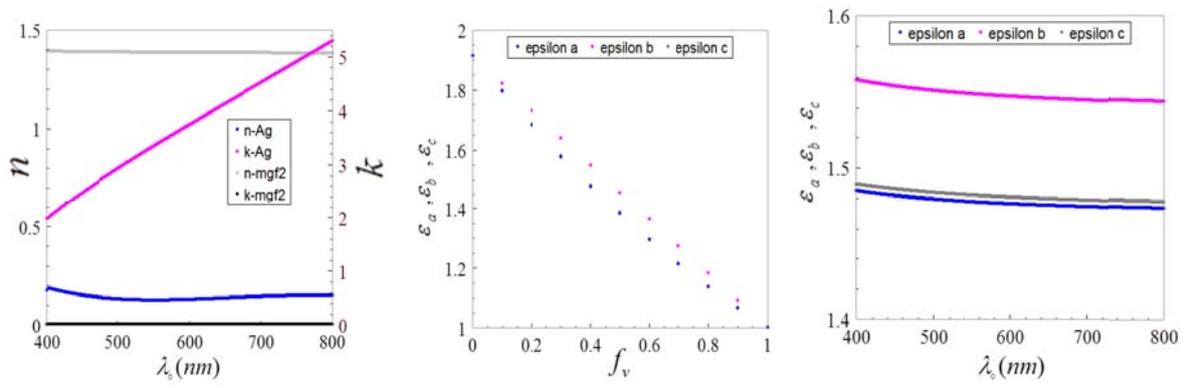

**Fig. 2;   F. Babaei and M. Omidi**

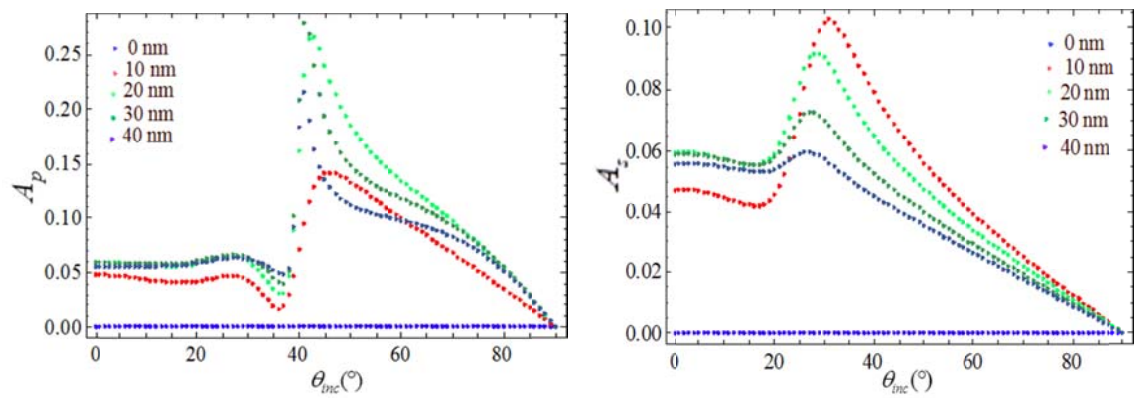

**Fig. 3;   F. Babaei and M. Omidi**

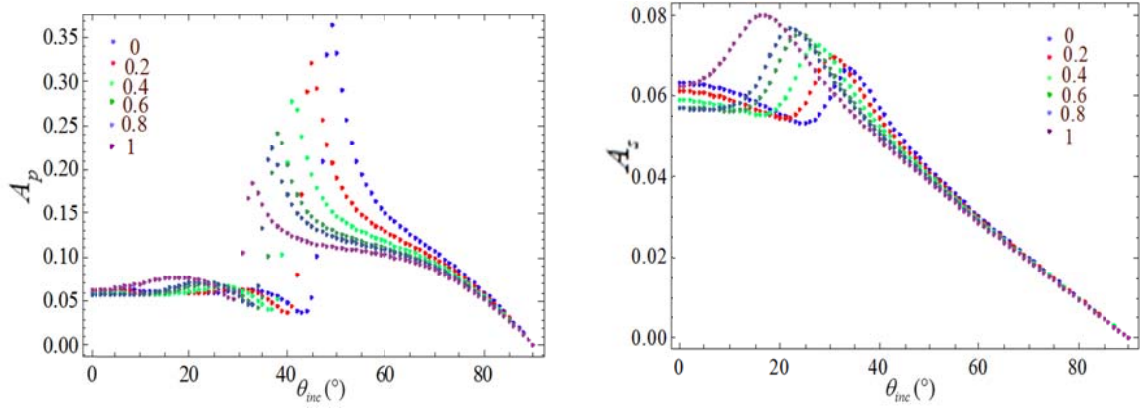

**Fig.4;   F. Babaei and M. Omidi**

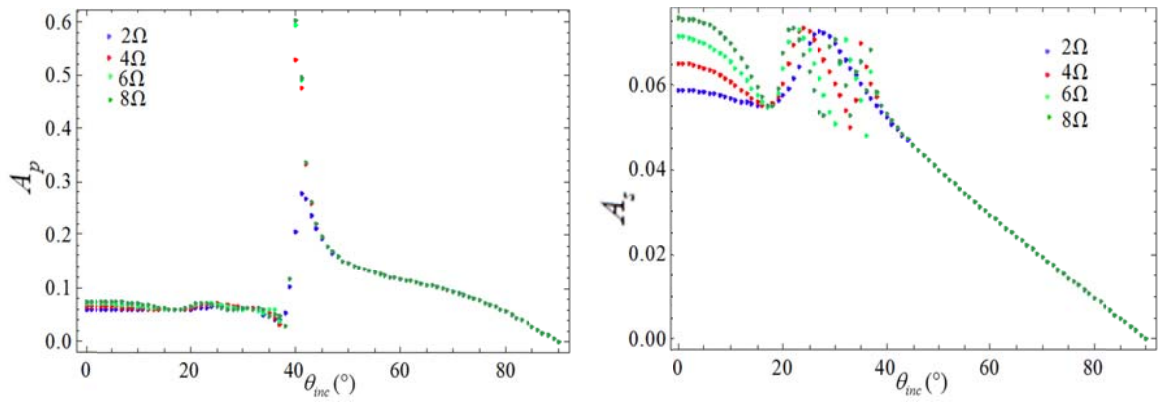

**Fig. 5;   F. Babaei and M. Omidi**

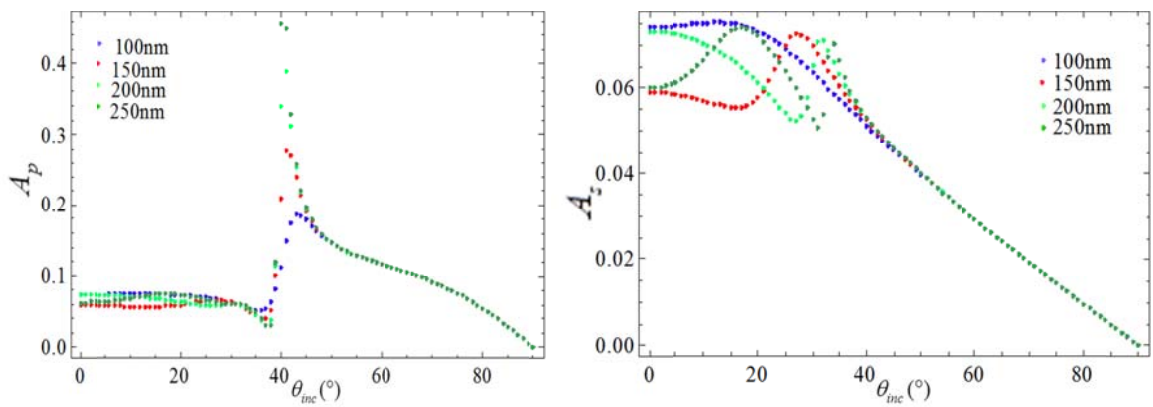

**Fig. 6;   F. Babaei and M. Omidi**

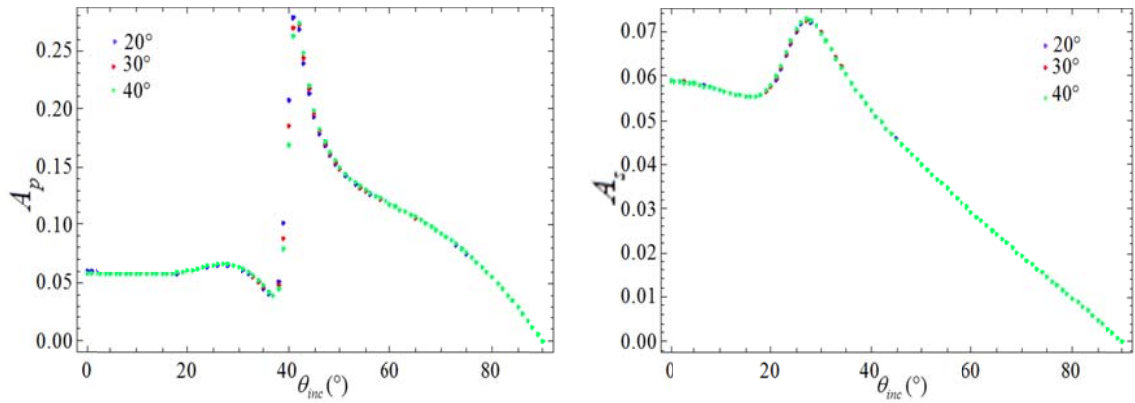

**Fig.7;   F. Babaei and M. Omidi**

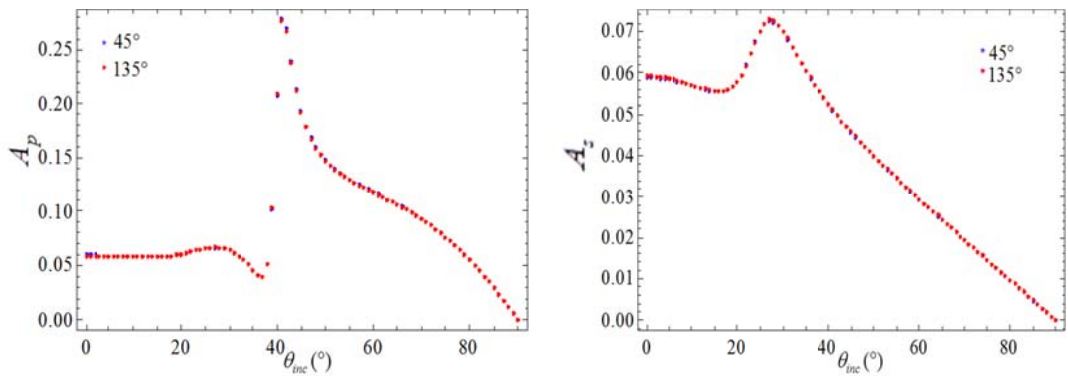

**Fig. 8;   F. Babaei and M. Omidi**

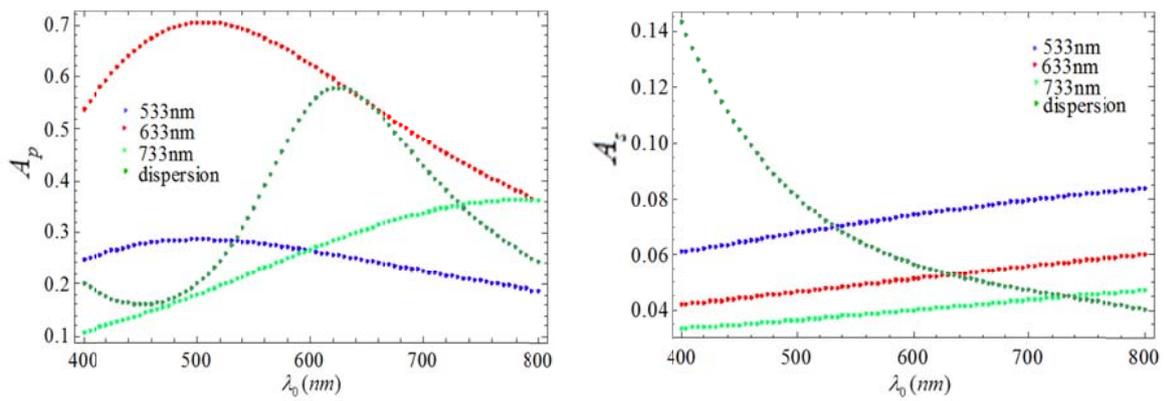

**Fig. 9;   F. Babaei and M. Omidi**

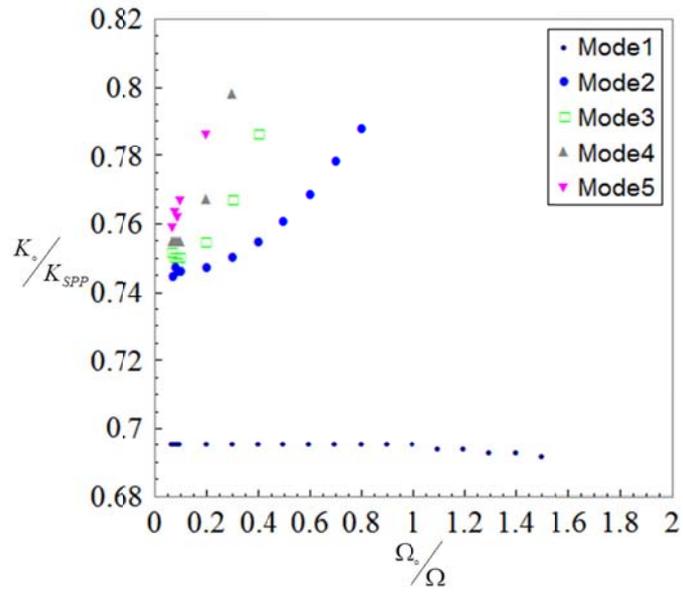

**Fig. 10;   F. Babaei and M. Omidi**

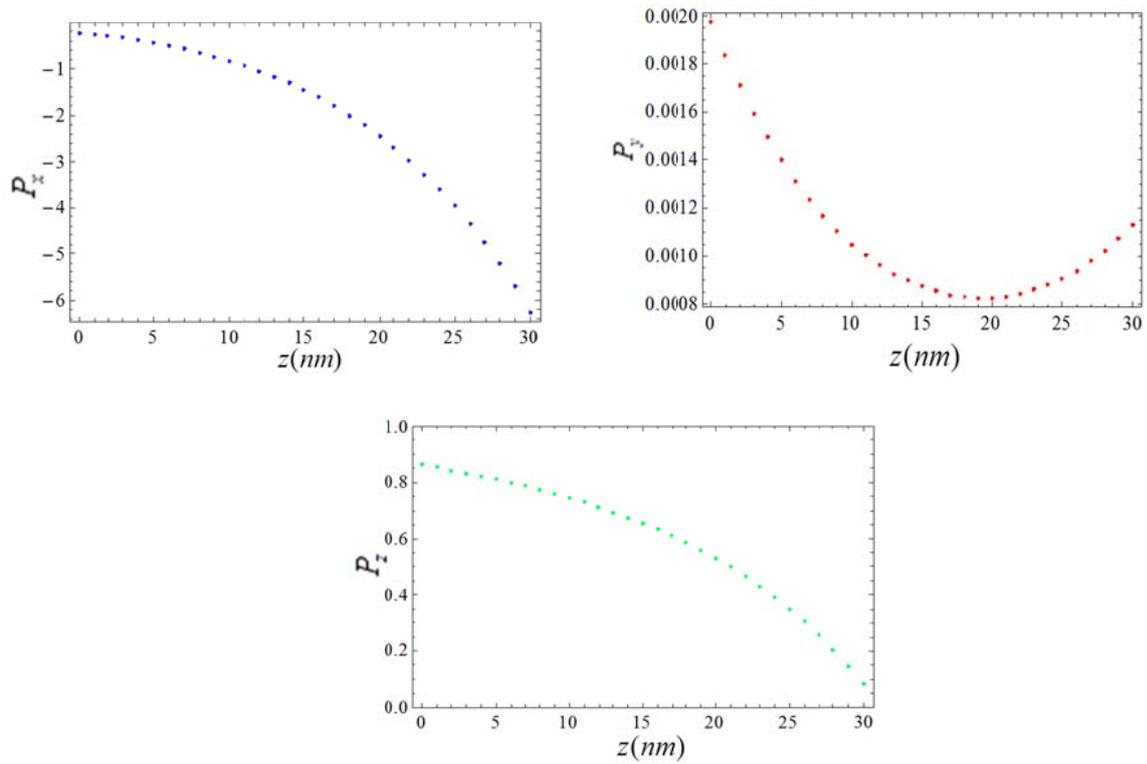

**Fig. 11;   F. Babaei and M. Omidi**